\documentclass[twocolumn]{article}

\usepackage{amsmath}
\usepackage{algorithmic}
\usepackage{array}
\usepackage{fixltx2e}
\usepackage{stfloats}
\usepackage{url}
\usepackage{xcolor}
\usepackage{fancyvrb}
\usepackage{graphicx}
\usepackage{comment}
\usepackage{multirow}
\usepackage{xspace}
\usepackage{subcaption}
\usepackage{fancyhdr}

\newcommand{\ant}{\texttt{Ant}\xspace}
\newcommand{\maven}{\texttt{Maven}\xspace}
\newcommand{\gradle}{\texttt{Gradle}\xspace}
\newcommand{\javac}{\texttt{javac}\xspace}
\newcommand{\jbf}{JBF\xspace}
\newcommand{\java}{\texttt{Java}\xspace}
\newcommand{\jar}{\texttt{JAR}\xspace}
\newcommand{\github}{GitHub\xspace}

\title{The Java Build Framework: Large Scale Compilation}

\author{Pedro Martins \ \ \ Rohan Achar \ \ \ Cristina V. Lopes  \\
    Institute for Software Research \\
	University of California, Irvine  \\
	\{pribeiro,rachar,lopes\}@uci.edu \\ \\ 
	UCI-ISR-18-3, April 2018
	}

\date{}

\begin{document}

\maketitle

\begin{abstract}
Large repositories of source code for research tend to limit their utility to static analysis of the code, as they give no guarantees on whether the projects are compilable, much less runnable in any way. The immediate consequence of the lack of large compilable and runnable datasets is that research that requires such properties does not generalize beyond small benchmarks.

We present the Java Build Framework, a method and tool capable of automatically compiling a large percentage of Java projects available in open source repositories like GitHub. Two elements are at the core: a very large repository of JAR files, and techniques of resolution of compilation faults and dependencies.
\end{abstract}

\newcommand{\projectCount}{353,709\xspace} 

\newcommand{\androidSuccessCount}{16,507\xspace}

\newcommand{\JBFsuccess}{190,727\xspace}
\newcommand{\JBFsuccessWithoutbuild}{103,801\xspace}
\newcommand{\JBFfail}{162,982\xspace}
\newcommand{\JBFfirstbuild}{92,482\xspace}
\newcommand{\JBFsecondbuild}{93,693\xspace}
\newcommand{\JBFthirdbuild}{4,552\xspace}
\newcommand{\JBFdependencyerrors}{164,546\xspace}
\newcommand{\JBFencodingerrors}{19,918\xspace}
\newcommand{\JBFsuccessWithbuildfile}{0\xspace} 

\newcommand{\projectsWithOwnFiles}{189,220\xspace}
\newcommand{\projectsWithMaven}{122,897\xspace}
\newcommand{\projectsWithGradle}{11,942\xspace}
\newcommand{\projectsWithAnt}{54,381\xspace}

\newcommand{\projectsWithMultiOwnFiles}{0\xspace}

\newcommand{\numberJars}{245,648\xspace}

\section{Introduction}

\java is one of the most popular programming languages, and the third
most-used programming language in
\github.~\footnote{\url{https://octoverse.github.com}, Dec.  2017} The
maturity and wide-spread acceptance that \java achieved justifies its
popularity among the research community, with considerable research
focusing on this language. It is therefore a prime candidate for
software mining. A particularly popular type of software mining
research is the extraction of source code features with the purpose of
performing various studies, such as correlation with bugs, security,
etc. These types of studies have motivated the development and
curation of repositories and datasets for large-scale source code
analysis, such as the Sourcerer
Datasets~\cite{Bajracharya:2009:SIS:1556907.1556984}, the
Qualitas Corpus~\cite{QualitasCorpus:APSEC:2010}, and
Boa~\cite{Dyer:2013:BLI:2486788.2486844}. 

These large repositories of source code for research, however, tend to
limit their utility to static analysis of the code, as they give no
guarantees on whether the projects are compilable, much less runnable
in any way. To the best of our knowledge, no public large-scale
repository of \java projects provides guarantees that the projects
compile, or run, something very relevant to a wide spectrum of
analysis techniques. Those kinds of guarantees can be made on small,
manually curated datasets, such as those of the
DaCapo~\cite{Blackburn:2006} and SPEC~\cite{SPEC} benchmarks.  The
immediate consequence of the lack of large compilable and runnable
datasets is that conclusions from research work that require such
properties (e.g.
\cite{Ernst:2000:QDR:337180.337240,Flanagan:2009:FEP:1542476.1542490,Joshi:2009:RDP:1542476.1542489})
may not generalize beyond the small benchmarks with which they have
been evaluated. The scope and impact of these studies would be much
greater if they could leverage the vast amounts of open source \java
programs available, similarly to what has already been achieved for
studies that require vocabulary, software metrics, and static
analysis.

But scaling up compilation, testing, and execution to thousands of
projects from Internet repositories is a daunting task. In order to
compile, test, and run thousands of projects, we need automated
techniques for resolving dependencies (compilation), for finding and
driving test suites (testing), and for producing input and workflows
(execution). These three types of automation are all challenging in
different ways, with the last one (execution) being the most difficult
one.

This paper presents a first step into obtaining such datasets by
tackling the first challenge: large-scale compilation. We present the
Java Build Framework (\jbf), a method and tool capable of automatically
compiling a large percentage of Java projects available in open source
repositories like \github. Two elements are at the core of \jbf: (1) a
very large repository of existing \jar files, and (2) a technique of
dependency resolution that identifies the external types in the
projects and maps them to the most appropriate \jar.

Along with \jbf, we provide a repository of 50,000 compilable (and
compiled) \java projects taken from \github, called 50K-C. Each
project in this dataset comes with references to all the dependencies
required to compile it, the resulting bytecode, and the scripts with
which the projects were built. The dependencies, bytecode, and scripts
are all products of \jbf. The dependencies and the build scripts
provide a mechanism to re-create compilation of the projects, if
needed (to instruct source code for bytecode analysis, for example). The bytecode is ready for testing, execution, and dynamic
analysis tools. All the 50,000 projects come with origin information,
which means that analyses on them can be cross-referenced with other
tools operating on \github , such as
Boa~\cite{Dyer:2013:BLI:2486788.2486844} or the map of \github clones
D\'{e}j\`{a}Vu\cite{Lopes:2017:DMC:3152284.3133908}.

One advantage of the 50K-C dataset with respect to many other large
datasets of source code is that all project dependencies have been
resolved by \jbf. This, in itself, is valuable for many research
studies, independent on whether execution is a goal or not. Another
advantage is that, even for static analysis, compilation gives solid
assurances that the source code is complete and self-contained: having
been parsed successfully means that that source code is valid Java
code; having been type checked successfully means that all required
types, external and internal, are present. For assurances of physical
integrity, compilation is the ultimate test.

While large-scale compilation is just the first step towards
developing very large collections of runnable projects for research,
it is a critical one, without which the other two steps are not
possible. Resolving dependencies of arbitrary projects taken from
sites like \github is a challenge in itself. This paper explains our
method of doing it, and compares it to the baseline of trying to use
the build scripts included in the projects, when they exist. It also shows
how current build techniques, despite sophistication and success on individual projects, struggle to support
the large scale buildings for which they were never designed.

The source code, datasets and running tutorials can be found on \url{http://mondego.ics.uci.edu/projects/jbf}.

This paper is organized as follows. Section~\ref{sect:jbf} gives an
overview of \jbf. Sections~\ref{sec:jars} and \ref{sec:compilation} present
the two main parts of the framework. Section~\ref{sec:eval} describe
the evaluation of \jbf. Section~\ref{sec:discussion} provides statistical observations
of build successes. Section~\ref{sec:relwork} presents
related work, and Section~\ref{sec:conclusion} concludes the paper.

\section{\jbf}\label{sect:jbf}

\begin{figure*}
  \centering
  \includegraphics[width=0.9\textwidth]{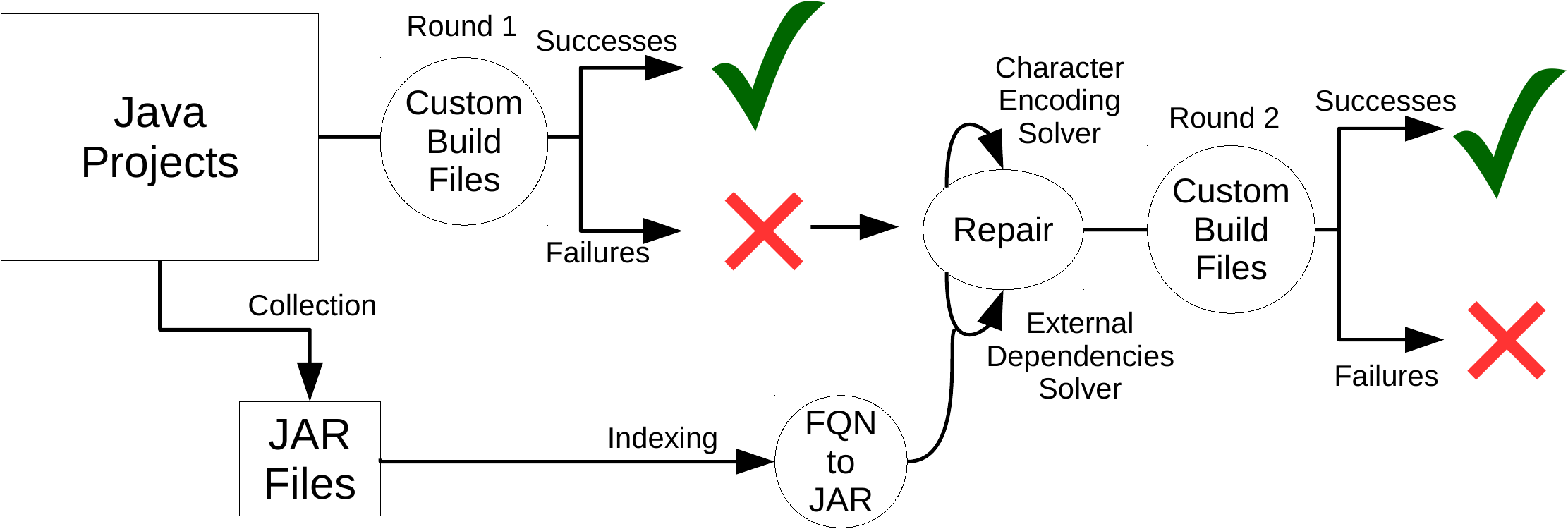}
  \caption{SourcererJBF pipeline.}
  \label{fig:sourcererjbf}
\end{figure*}

An overview of the entire process that constitutes \jbf can be seen in
Figure \ref{fig:sourcererjbf}. \jbf is split into two main tasks, namely
management of dependencies and project compilation, and each of these
is further split into sub-tasks.

Starting with dependency management, the first task on the pipeline
consists of collecting \jar files from the projects. After that, \jbf
creates an inverted index of Fully Qualified Names (FQN) of \java
classes and interfaces available in a \jar. This is done before any
compilation takes place.

The compilation process in \jbf is done in up to two rounds for each
project. In a first round, an attempt is made to compile it without
specifying any external dependencies, just relying on the \java
standard libraries. The ones that succeed are marked as so; the ones
that fail go through a second round of processing.  \jbf detects
compilation errors related to encoding problems and to missing external
dependencies. Both are solved through inserting specific directives in
our compilation scripts. For resolving external dependencies, we use
the index created in the initial phase.
A second round of compilation starts whose results are final: the
projects that compile are marked as successes, the failures as
failures, and the results of this round together with the results of
the first round determine the overall effectiveness of \jbf.

In the next sections we will explain these tasks of \jbf in detail.

\section{Collection and Management of Dependencies}
\label{sec:jars}

Reliance on external dependencies is nowadays typical of software
systems. Therefore, our first concern when attempting to build a set
of software projects is to ensure availability of dependencies in
order to minimize the number of failed builds.

Generally, \jar dependencies required for compilation come from three main sources:

\begin{itemize}

\item Boot libraries, with generic functionality provided by the Java Virtual Machine runtime. These dependencies are automatically available to building tools.

\item Installed extensions through the Extension Mechanism Architecture\footnote{\url{http://docs.oracle.com/javase/6/docs/technotes/guides/extensions/spec.html}}, which are similar to the core libraries in usability but which vary from environment to environment. The dependencies available through this location extend core functionality and can have a varied origin, but must follow a strict set of guidelines.

\item External libraries explicitly referenced in the source code, and that can have any location on the system as long as the building tools are aware of their existence. These dependencies can have any origin. A typical case for the usage of these dependencies is a project that is stored together with a certain \jar file that it requires.

\end{itemize}

\subsection{Collection}

For the boot libraries we do not interfere with their typical usage, we simply expect any system under our operational pipeline to contain a standard \java installation.

For the extensions, we opt to avoid this mechanism entirely, by
directing the compiler \javac to an empty location, overriding
the system's default behavior. The reason for this is environment
control: these libraries change between different \java versions, with
JDK updates within the same \java version (\java 8.1 to \java 8.2) or
with personal preferences in different installations of the same
versions of the JDK. Instead, we place a handful of commonly used
extension libraries along the external libraries.

For the external libraries, we scrape all the projects we have for
\jar files that were packaged within them, and copy them to a central
repository, which consists of a folder with all the \jar files inside
(to be precise, a set of folders, because one single folder does not
scale). 

\subsection{Filtering}

We observed that some \jar files have invalid signatures that cause \javac to crash silently. 
We use the tool \texttt{jarsigner}\footnote{\url{http://docs.oracle.com/javase/7/docs/technotes/tools/windows/jarsigner.html}} to analyze every \jar we collect. In particular, we are only concerned with \jar files that are unusable due to some problem in their digital signature, for which \texttt{jarsigner} produces the following output:

\begin{Verbatim}[fontsize=\small]
java.lang.SecurityException: Invalid
signature file digest for Manifest main
attributes
\end{Verbatim}

The \jar files that produce this error are dismissed. All the others
are accepted into our central repository.

\subsection{Indexing}

Types in \java are referenced by their Fully Qualified Name (FQN). A
FQN is an identifier of a class or an interface which includes the
package where it belongs.\footnote{For a detailed explanation of FQN
  and general naming conventions in \java please see
  \url{https://docs.oracle.com/javase/tutorial/java/package/namingpkgs.html}.}
In order to assist dependency resolution, we create an inverted index
that maps every type we find inside \jar files to the \jar files where
it occurs.

Specifically, we analyze every \jar file in order to list its table of
contents. For example, consider a certain \texttt{uci-irv.jar} with the
following contents:

\begin{Verbatim}[fontsize=\small]
META-INF/MANIFEST.MF
edu/uci/ics/algo.class
edu/uci/econ.class
\end{Verbatim}

\noindent
This lists two types, each identified by an FQN. Upon index creation,
both FQNs are entered as keys, having this specific \jar file in their
list of values, generating the following index: 

\begin{Verbatim}[fontsize=\small]
edu.uci.ics.algo: {uci-irv.jar}
edu.uci.econ: {uci-irv.jar}
\end{Verbatim}

Returning to the usage of FQN to resolve types, \java allows
resolution through partial FQN to an entire group rather than to a
specific type. For example, source code that requires the \jar above
can specify this relation as importing \texttt{edu.uci.*}, with the
wildcard meaning that the requirement is to any class that is
hierarchically below \texttt{edu.uci}, which can be either
\texttt{ics.algo} or \texttt{econ.class}, or both. This
results in ambiguity that can be solved in a couple of different
ways. For performance reasons, we index all possible FQNs that can be
used in a project. Returning to our running example, the following FQNs
will be indexed, all pointing at the same \jar file:\footnote{We do
  not index \texttt{org} or \texttt{com}, for obvious reasons: no project imports
  \texttt{com.*} or \texttt{org.*}.}

\begin{Verbatim}[fontsize=\small]
edu.uci.ics.algo: {uci-irv.jar}
edu.uci.econ: {uci-irv.jar}
edu.uci.ics: {uci-irv.jar}
edu.uci: {uci-irv.jar}
\end{Verbatim}

Note that the mapping between FQN and \jar files is one-to-many, since
two \jar files may contain the same FQN. For example, if another \jar
had the FQN \texttt{edu.uci.psico}, then it would be
indexed in same positions where
\texttt{edu.uci.psico} (and its reduced variations)
intersect with \texttt{uci-irv.jar}, namely 
\texttt{edu.uci}.

The final result of indexing is a large mapping between all FQNs found
and the paths of the respective \jar files.

\section{Projects Compilation}
\label{sec:compilation}

After having the external dependencies organized as an inverted index,
we can start building the projects. This process can take up to three phases: a
first round of compilation, a middle task of error repair, and a
second round of compilation.

\subsection{Building Projects - Round 1}\label{subsect:round1}

The first round of compilation does not rely on external dependencies
and has two objectives: first, to immediately build the projects whose
compilation is straightforward; and second, to collect a list of
errors for projects that fail and provide \jbf with information to fix
them.

For every project \jbf creates a generic \ant build file, which simply
invokes \javac for all java files found from the project's root folder
and subfolders. After this call, two things can happen: either the project compiles, or it fails.
If the project compiles, we store it as a success. For each project that \jbf builds we store:

\begin{itemize}

\item the resulting Class files;

\item the generic build files (they are generic but contain some information related to the file system, so we keep them);

\item the exact system call that successfully called the build scripts and built the project;

\item and the output produced when building the project.

\end{itemize}

The projects that fail are passed to the error repair mechanism of \jbf.

\subsection{Error Repair}

For the projects that fail after Round 1, a new task starts, where \jbf parses the resulting output of \javac and searches both for character encoding errors and for missing packages. These type of errors take the forms of

\begin{Verbatim}[fontsize=\small]
error: unmappable character for encoding UTF8
\end{Verbatim}

\noindent
for encoding errors, and

\begin{Verbatim}[fontsize=\small]
error: package org.msr does not exist
\end{Verbatim}

\noindent
for errors related to missing external dependencies.

\jbf parses the output of \javac looking for signs of these errors. For encoding errors, \jbf runs an analysis of the files and passes specific encoding information to \javac, such as:

\begin{Verbatim}[fontsize=\small]
encoding="windows-1252"
\end{Verbatim}

For errors related to external dependencies, \javac reports them in
the form of missing FQNs. We then use them as queries to the index to
find the \jar files in which they exist. 

The dependency solver of \jbf captures all the missing packages from
an unsuccessful build and finds the \jar file that contains the
largest number of them. For example, if the failed attempt of the
project reports 10 missing packages, and we find a \jar file that
contains a reference to all of them, then this is the \jar file
chosen. If we find a \jar file that contains a reference to 8 of these
missing packages, then we choose it, and recursively start the process
for the next two. If more than one \jar file matches a
certain requirement (for example, more than one \jar file contains a
reference to all of the 10 missing packages), then we choose the first
one without any particular preference.

In the end, all the \jar files that are found are considered to be the
necessary external dependencies, and are given as such to \javac. 

An example of the code that is inserted into \javac task of \ant to fix dependencies is:

\begin{Verbatim}[fontsize=\small]
<pathelement path="path/to/uci-irv.jar" />
\end{Verbatim}

An important note is that, in this stage, no FQN reduction is
performed when trying to resolve dependencies, for a simple reason:
when we are scraping and indexing the \jar files, we have access to
all the complete FQNs that the \jar file contains, that is,
all the FQNs from the top of the package hierarchy until a certain
Class. On the project side, the \texttt{package
  not found} error depends on what the programmer specifically wrote
in the source code. For example, the programmer might have specified a
requirement for \texttt{edu.uci.*} when she is using the Class
\texttt{edu/uci/ics/algo.class}. For this reason, if we can not
find on the indexed \jar files a full FQN from a package, we are
certain that Class does not exist. On the other side, if the reduction
comes from the project side then we can attempt to find a \jar file
that matches that reduced format, and hope it will contain the
required Class. This is the reason why we indexed all the \jar files
by all the possible FQNs that can be used to refer to them.

If the dependency solver can not find a \jar file, it throws a
\texttt{Missing package error}.

\subsection{Building Projects - Round 2}

The second round of compilation consists on building the project that
failed on the first round, with specific directives that try to
resolve both the encoding and the dependency errors as described above.

After this round, the projects that build are stored with the same
information as described before, now with build scripts that contain
encoding and dependency information that is specific to the actual
project.

The projects that fail after this step are considered true fails, and
no new attempt is made to try to build them again. Their build
information is stored nevertheless, in a format that is similar to the
one for successful buildings, except in the output we write the
errors the \javac compiler produced, and there is no folder containing
Class files, because no output was generated.

\section{Evaluation}
\label{sec:eval}

Here we present the evaluation of \jbf, first alone, then in
comparison with the build scripts that come with the projects, for the
projects that contain them. Our concrete research questions driving
this evaluation are the following:

\begin{itemize}
\item What is the effectiveness of \jbf on building \java projects on scale?
\item How does JBF compare to the projects' own build scripts, when present?
\end{itemize}

\subsection{Testing Corpus}\label{sect:corpous}

To test \jbf we downloaded a set of \java projects from \github. The projects were downloaded using the GHTorrent database and network\footnote{\url{http://ghtorrent.org}}~\cite{Gousi13} which
contains meta-data from \github (number of stars, commits, commiters, whether the projects are forks, main programming language, date of creation, etc.) as well as the download
links.\footnote{While GHTorrent enabled us to crawl large amounts of repositories, we have found its database to have some errors. Specifically, a residual number of duplicated entries 
(i.e. projects had the same URL); only the youngest of these was kept for the analysis. We filed an issue report at (URL omitted for anonymity).}
  
\begin{table}[!ht]
\centering
\caption{Corpus of Java projects.}
\label{tab:dataset}
\begin{tabular}{|l || r |} \hline
 \# projects (total)    & 3,506,219\\ 
 \# projects (non-fork) & 1,859,001\\
 \# URLs processed & 631,390 \\
 \# projects (downloaded) & 479,113\\
 \# projects (excluding Android) & \projectCount\\ 
 \# jars & \numberJars\\
 \# FQNs & 8,164,106 \\\hline
\end{tabular}
\end{table}

\begin{figure}
  \centering
  \includegraphics[scale = .5]{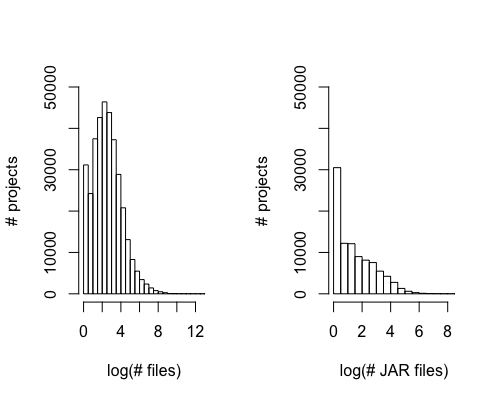}
  \caption{A distribution of the number of files and number of \jar files in the \java projects.}
  \label{fig:project-stats-files-jars}
\end{figure}

Table \ref{tab:dataset} shows the information regarding the size of
the language corpora in our analysis. We skipped forked projects,
because since they represent directly cloned projects and usually
contain a large amount of code from the original projects, having
explicit forks in the dataset for this study could skew the
findings. Figure \ref{fig:project-stats-files-jars} shows the
distribution of files and \jar files among the projects.

Downloading the projects was time-consuming. The order of downloaded
projects was that of the GHTorrent database projects table, which
seems to be, roughly, but not strictly, chronological.\footnote{The
  documentation of GHTorrent does not specify the exact order, if any,
  by which projects are added to the projects table. In analyzing the
  data, we noticed that although the projects are not ordered in
  strict chronological order, there seems to be a concentration of
  older projects in the earlier entries of the table. We filed another
  issue report asking the developers of GHTorrent to clarify this.}
Even though for \java we processed only 34\% of the available,
non-forked project URLs, the portion we processed provides a large
enough corpus to be significant.

Some of the URLs, when downloaded, failed to produce valid
content. This happened in two situations: (1) when the projects had
been deleted from \github, or marked private, and (2) when development
for the project happens in branches other than the master branch. As
such, the number of downloaded projects was considerably less than the
number of URLs available from GHTorrent. Some projects did not have
any source code with the expected extension, and these were excluded
from the analysis.

Finally, we also filtered the projects that represent software for the
Android ecosystem, since these have special requirements (we would
need, for example, to have different versions of the Android JDK.). We
did so by searching inside the projects for a
\texttt{AndroidManifest.xml} file, since according the respective
documentation, these files are mandatory for every Android
application.\footnote{\url{https://developer.android.com/guide/topics/manifest/manifest-intro.html},
  May 2017.}

All the tests were performed on a Intel(R) Xeon(R) CPU E5-4650 v4 @ 2.20GHz with 112 cores and 250 Gb of RAM.
\subsection{\jbf}\label{sect:jbf-performance}

In this section we present the results of running \jbf on the corpus of \projectCount \java projects. On total, between the two rounds of compilation \jbf compiled more than half the projects, \JBFsuccess, which corresponds to 54\%.

\begin{figure}
  \centering
  \includegraphics[scale = .5]{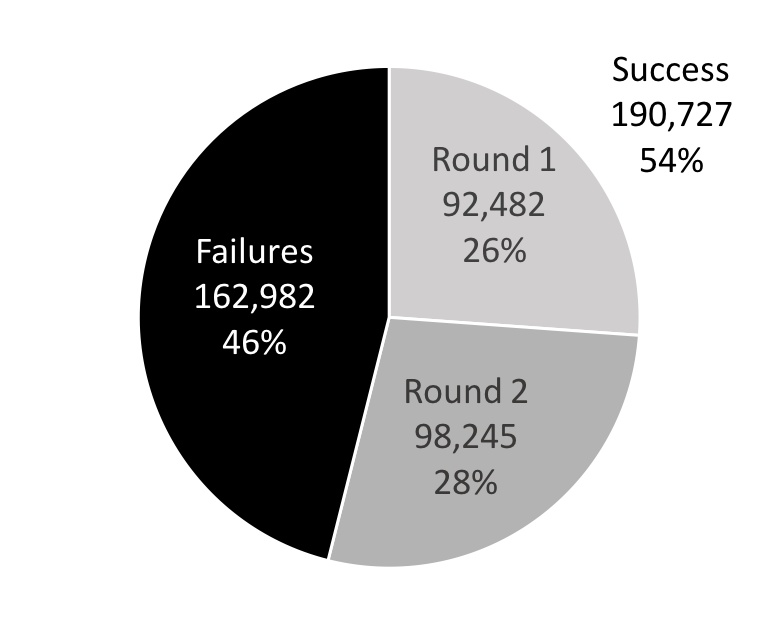}
  \caption{Number of projects built by \jbf.}
  \label{fig:JBF_success}
\end{figure}

From all the projects that \jbf was capable of building, around half built after the first round of compilation, meaning they did not require any kind of special assistance. The other half required the assistance of the error repair mechanism, which proved successful for 98,245 projects. The exact values of the number of projects built after the two rounds can be seen in Figure \ref{fig:JBF_success}.

\begin{figure}
  \centering
  \includegraphics[scale = .5]{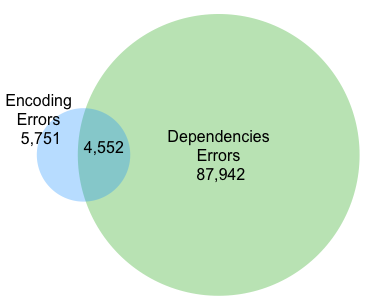}
  \caption{Distribution of errors handled per project built by \jbf.}
  \label{fig:JBF_errors}
\end{figure}

Regarding the work performed by the error repair mechanism we can see in Figure \ref{fig:JBF_errors} the distribution of errors among the projects that successfully built only after the second run, i.e., the distribution of errors for which \jbf behaved successfully. \jbf solved encoding errors in 10,303 projects, and resolved the external dependencies of 92,494 projects. 

For the projects that failed after the second round of compilation, which amounts to \JBFfail projects, we can see on Table \ref{table:top20errors} the most common errors. An error in special deserves attention, \texttt{Missing packages}, which was produced by the type-resolution mechanism, and means that no \jar file could be found for the specific type needs of a project. The other errors on the table were extracted directly from the output of the \javac compiler.

A broad analysis on Table \ref{table:top20errors} leads us to notice that most errors are related to missing types. This indicates both a lack of soundness on the type structure of projects, and/or a reliance on external type information for completeness. The conclusion is that any system that attempts to build source code on scale should be prepared to face some form of type problems, an observation we have done before, and that partly motivates this work, and gains here further support.

\begin{table}[]
\centering
\caption{Top 20 errors for failed projects. Note that these groups intersect, and most projects had more than 1 error type.}
\label{table:top20errors}
\begin{tabular}{| l || r |} \hline
Error & Frequency \\ \hline \hline
cannot find symbol & 210,949 \\ \hline
package not found & 183,614 \\ \hline
Missing packages & 124,742 \\ \hline
method does not override or implement & \\ 
a method from a supertype & 68,586 \\ \hline
duplicate class & 26440 \\ \hline
static import only from \\ classes and interfaces & 16,703 \\ \hline
unmappable character & 15,988 \\ \hline
expected symbol not found & 13,904 \\ \hline
illegal access & 13804 \\ \hline
Too many encoding types detected & 12,220 \\ \hline
incompatible types & 12,155 \\ \hline
illegal use & 9,367 \\ \hline
cannot be applied & 7,469 \\ \hline
no suitable definition found & 6,122 \\ \hline
class should be in its own file. & 5,541 \\ \hline
abstraction error & 5,329 \\ \hline
not a statement & 3,541 \\ \hline
reached end of file while parsing & 3,081 \\ \hline
invalid method declaration; \\ return type required & 1,816 \\ \hline
too many parameters & 1,631 \\ \hline
\end{tabular}
\end{table}

As a side note, \jbf successfully built \androidSuccessCount projects that were specifically Android. These projects were not added to the statistics presented until this point, and they are not part of the evaluation corpus we have presented in Section \ref{sect:corpous}, since we removed Android projects. We are adding this information here just to note that the general approach of using generic build systems had some success in these cases.

\subsection{Build Frameworks}

We now compare the effectiveness of the build frameworks stored with the projects with the efficiency of \jbf. 

The build frameworks within the projects were detected by searching for specific files that would flag their existence. In particular, we searched for a \texttt{build.xml} file to detect the existence of \ant, a \texttt{pom.xml} file to detect the existence of \maven and a \texttt{build.gradle} file to detect the existence of \gradle.

\begin{figure}
  \centering
  \includegraphics[scale = .5]{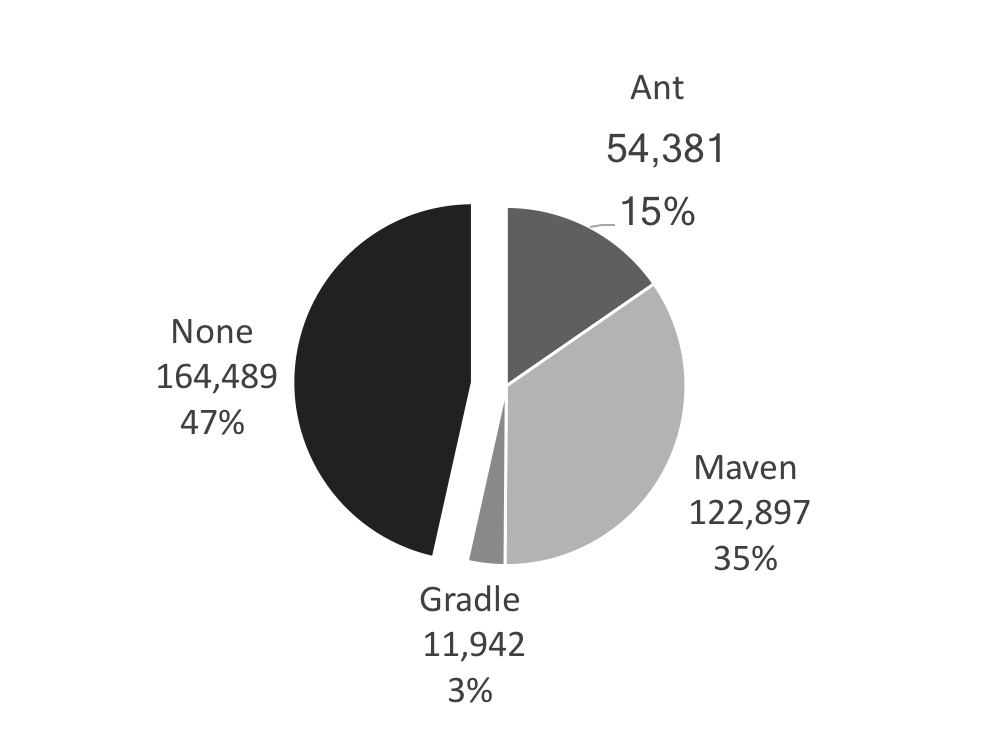}
  \caption{Distribution of build systems across the projects.}
  \label{fig:build-systems-dist}
\end{figure}

We present, on Figure \ref{fig:build-systems-dist}, the distribution of build frameworks across the entire corpus of \java projects. On total, \projectsWithOwnFiles projects (53\% of the \projectCount total projects) had build information of some sort. The first important observation is that around half the projects exists without information on how to compile them. This might be due to their nature (not all projects represent a single software solution), or due to the nature of their origin, either for lack of proficiency or personal preference.

Operationally, the test of the build systems was instructed to directly call whichever build framework existed with the projects. In cases where there was more than one, the order of prioritization was \ant first, then \maven and then \gradle (the number of projects in this situation was residual). For each of the projects with a build framework we made a generic system call that would start the process of building it. For \ant, \maven and \gradle the calls were, respectively:

\begin{itemize}

\item \texttt{ant -f <path to build file>}

\item \texttt{mvn -f <path to pom file> compile}

\item \texttt{gradle -b <path to gradle file> compileJava}

\end{itemize}

Only one try was allowed per project, without any type of error resolution: either it immediately succeeded, or immediately failed.

\begin{figure}
  \centering
  \includegraphics[scale = .5]{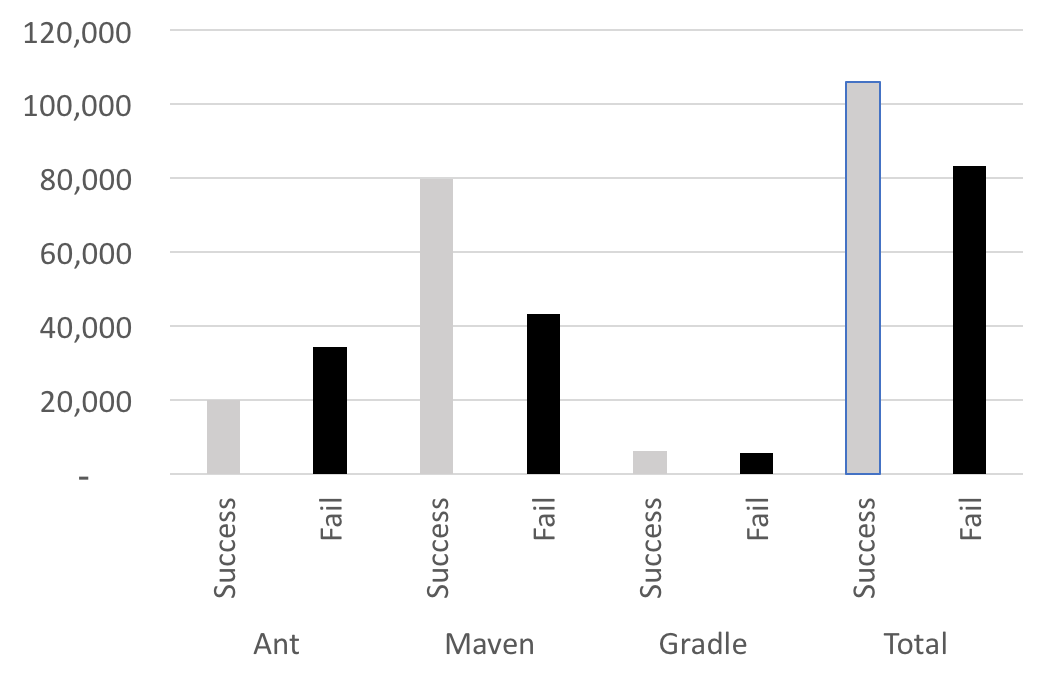}
  \caption{Compilation success rates for each build system individually.}
  \label{fig:builds-distribution}
\end{figure}

We start by observing, in Figure \ref{fig:builds-distribution}, the effectiveness of the build frameworks in isolation as well as their aggregated performance. \ant is the build framework that performs the worst, and is only capable of building 20,095 projects, failing in the other 34,286 projects. \maven is the most successful, and builds more than half the projects; on total, \maven is capable of building 79,672 \java projects, and fails only for 43,225. We have seen before that type errors, in particular missing external types are frequent. Knowing that \maven has the capacity of automatically managing external dependencies, namely obtaining them from online sources, provides a likely explanation as to why this tool performed so well.
\gradle successfully built 6,206 \java projects and failed for 5,736, splitting more or less in half the success rate of this tool.

\begin{figure*}
  \centering
  \includegraphics[scale = .5]{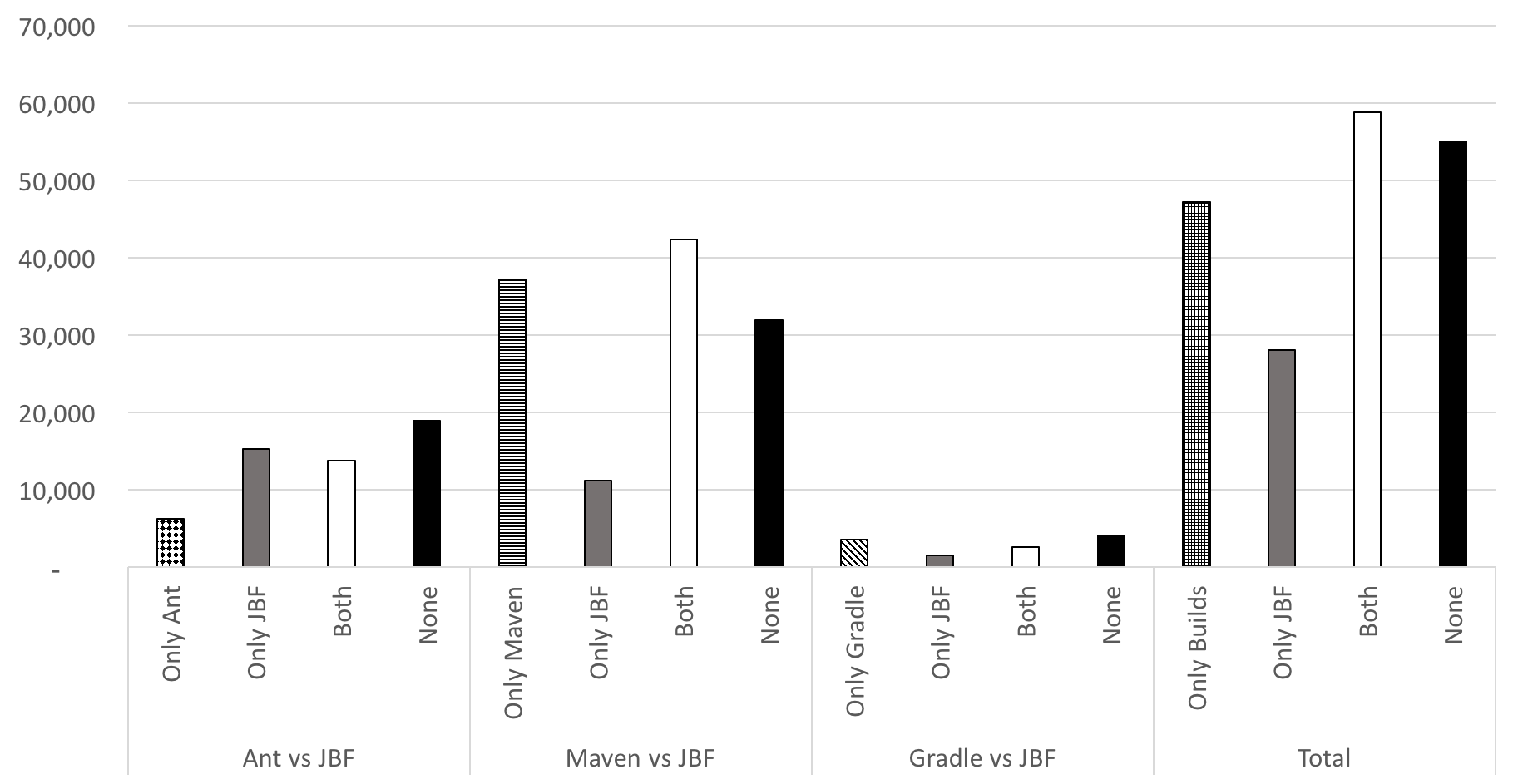}
  \caption{All build systems vs. \jbf, \projectsWithOwnFiles projects total.}
  \label{fig:jbv-vs-builds-comparison}
\end{figure*}

Going now to the comparison between \jbf and the build systems, in Figure \ref{fig:jbv-vs-builds-comparison} we can see individual and aggregated comparisons with \jbf. Starting with \ant (first chart), \jbf performed better, managing to build 15,332 of the projects the tool was not capable to, while the inverse is only true for 6,321 projects. Both \jbf and \ant successfully built 13,774, and both failed building 18,954 projects.

\maven was the framework that performed better in comparison to \jbf, managing to build 37,252 projects where \jbf failed, whereas \jbf built only 11,223 of the projects in which \maven failed. For \maven we have a large intersection of projects built by both tools, 42,420, which means that both strategies performed with a reasonable degree of success on the group of \maven projects.

The results for \gradle (third chart) are similar to \maven. Both strategies successfully built around a quarter of the projects, and \gradle performed better than \jbf but, in comparison, not as good as \maven.

Finally, we can analyze the aggregate of the building tools against \jbf, in the fourth chart of Figure \ref{fig:jbv-vs-builds-comparison}. Both the build frameworks and \jbf successfully built 58,785 projects, and the build frameworks built exclusively 47,188 projects, being the \jbf counter-part 28,141 projects. Contextualizing with all the projects, the successes and failures shared by \jbf and the build frameworks cover around two thirds of all the projects, for the other third, the build frameworks perform more effectively.

\subsection{Build Frameworks and Scale}

There are a few notes worth mentioning for understanding the performance of build frameworks on environments where they have to scale. 

The first aspect is that when comparing \jbf with existing build systems we are inherently neglecting a large number of projects that simply do not contain one. Within all the \projectCount projects we obtained, if we used build files we would be able to build 105,973 projects, whereas with \jbf we were capable of building \JBFsuccess projects, out of which \JBFsuccessWithoutbuild projects contain no build information.

Second, the build frameworks analyzed are in their essence scripts that allow a possible infinite number of actions on a large number of domains. They can remove, add or move information in the system or connect to online servers. It is easy to see the security holes that running a large number of unknown script creates, and these should not be taken lightly.

Third, the utility of building software is in the access of the produced runnable elements, in our case class files. But the use of unknown scripts can - and often does - cripple this advantage. We do not know where the targets are being allocated, or if the script is creating a JAR file instead of class files.

Fourth, the whole notion of success is dependent only on the full execution of the build framework, not on its intrinsic actions. This means a script can be successful but everything it does is printing the famous 'Hello World!'. On efforts of building on scale this introduces an uncertainty that easily becomes unreasonable. For example, when we provide the results of the build scripts in the previous sections these represent successes from the perspective of the operating system (the process terminated without an error code), not from the perspective of the inherent compilation task. This problem is of hard leverage.

\begin{figure}
    \centering
    \begin{subfigure}[b]{0.5\textwidth}
        \includegraphics[width=\textwidth]{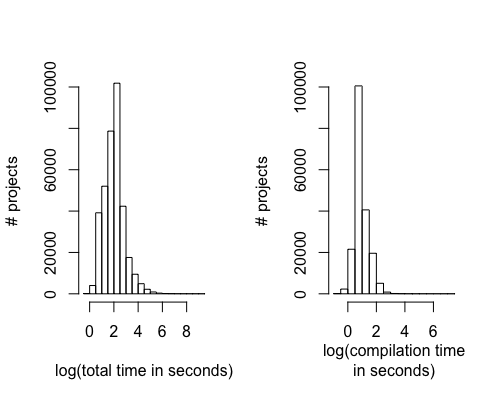}
        \caption{\jbf.}
        \label{fig:gull}
    \end{subfigure}

    \begin{subfigure}[b]{0.5\textwidth}
        \includegraphics[width=\textwidth]{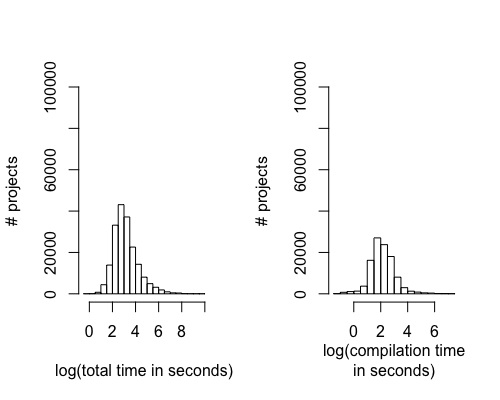}
        \caption{Build frameworks.}
        \label{fig:tiger}
    \end{subfigure}
    \caption{Time used to completely handle each project, and time used to compile each project using their building information (only for the ones that compiled successfully).}
    \label{fig:times}
\end{figure}

Fifth, and related with the previous two, the time it takes to run a build script is as much of a mystery as its contents. The building of the \projectCount \java projects using \jbf required a median of 7.5 seconds per thread, where each thread was responsible for the complete process from the management and extraction of the archived files to the compilation and any necessary clean-ups or error recoveries (Figure \ref{fig:times}). The actual compile time (a sub-task of each thread) had per project a median of 2.3 seconds. The equivalent times for the building frameworks were 20 seconds in median, and just to compile 7.8 seconds in median. The times, especially the compile time, are substantially longer than \jbf times. On universes of hundreds of thousands of projects this is relevant as it represents differences of weeks between one strategy or the other.

During our experiments running the build systems we found a little of all of this: random files appearing in the file system, 'stuck processes', lost target files, and so on. It is important to emphasize that this is not a critic of the build systems themselves: \ant , \maven and \gradle . These are quite sophisticated, very capable, and on single entities potentially more capable than \jbf . The problem is that they do not scale, they do not possess the characteristics of control, predictability or security that \jbf does simply because they were never designed with that concern in the first place.

\section{Analyzes of Success Builds}
\label{sec:discussion}

While executing and testing \jbf we realized some characteristics of \java source code are more likely to result in success builds than others. This provides statistic indicators for successful compilations which adds to the understanding of source code and can help in sample selection, relevant even on scale by limited resources (for example, which projects to download from GitHub to maximize build performance). For this reason we provide them here.

\begin{figure}
  \centering
  \includegraphics[scale = .5]{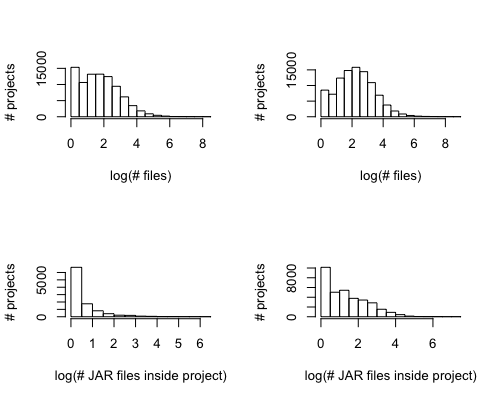}
  \caption{A comparison of projects that did not require error resolution (left) and projects that did (right). On top we have the distribution of the number of files, and on the bottom we have the number of \jar files that the projects contained.}
  \label{fig:firstvssecond}
\end{figure}

Between the two building rounds that exist in the \jbf pipeline, we have seen that each one successfully built around half the total successes of \jbf. Figure \ref{fig:firstvssecond} shows a distribution of the number of files and inside \jar files for the group of projects that was successfully built after the first round (left) and after the second round (right). It is noticeable that the first group has a smaller number of files (top-left) which leads us to speculate that the reason some projects build immediately is due to their simpler structure, at least if complexity of a project is measured in relation to the number of files it has. The distribution of \jar files for the projects that required some sort of error resolution (which, as we saw, was mostly related to external dependencies) have a larger number of \jar files inside them comparing to the ones that did not.

\begin{figure}
  \centering
  \includegraphics[scale = .5]{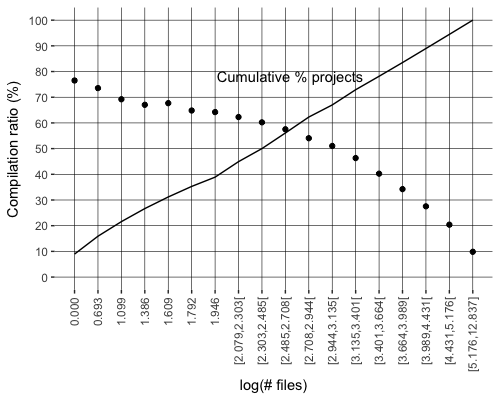}
  \caption{\jbf success ratios per project size.}
  \label{fig:files-vs-success}
\end{figure}

\begin{figure}
  \centering
  \includegraphics[scale = .4]{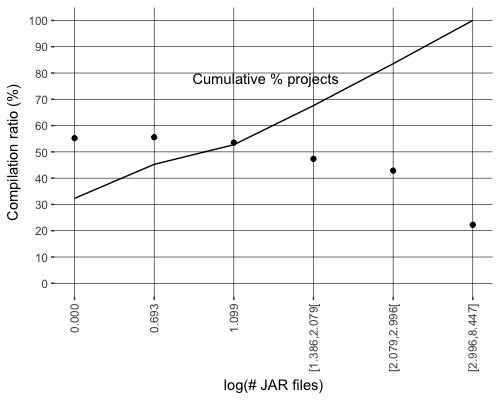}
  \caption{\jbf success ratios per number of \jar files inside the project.}
  \label{fig:jars-vs-success}
\end{figure}

For the relation of the number of files of a project with its compilation ratio we can look at Figure \ref{fig:files-vs-success}. The black dots are small intervals on the scale of the x-axis, representing subsets of similar sized of projects with different number of files ('bins'). The y-axis represents the compilation ratio; a certain, random value of 70\% implies that whichever bin (or black dot, or interval of size) is within it's range, it contains a subset of projects where 70\% successfully compile. 

The line labeled 'Cumulative \% projects' is a visual indication of how well the partition of bins splits the number of projects (the y-axis, showing percentage, has the same unit for compilation ratio and cumulative \% of projects). The line has an almost perfect growth, showing a good distribution of projects across the x-axis.

If we look at the first, highest point of the figure we see that its x-value is close to 0, which means this point represents projects that are very small. The y-value of the same point is around 75\%. This means that for projects that are very small, 75\% compile and 25\% fail. If we now look at the right-most point, its y-value is 10\%, so the subset of projects that are on the largest side of size have a small compilation ratios.

Observing the entire of the x-axis, and comparing to the evolution of the values on the y-axis, we can clearly see a trend of lowering compilation ratios as the size of the projects increases. For small projects we start with a compilation success ratio of around 75\%, and as we progress towards larger projects the ratio at which they compile is successively lower. The broad observation is that the number of files affects negatively the likelihood of a project compiling. The actual Pearson correlation coefficient is -0.95.

On Figure \ref{fig:jars-vs-success} we can see a similar assessment of the impact of the number of enclosed \jar files under compilation, which is negative. The impact on compilation is negative as the number of \jar files within the project grows. For projects with a very small number of dependencies the ratio of compilation is around 55\%, and as the number \jar files increases we see the compilation ratio decreasing to around 21\%. We found a correlation between the number of enclosed \jar files and the compilation success ratios, with a Pearson correlation coefficient is -0.88. 

\section{Related Work}
\label{sec:relwork}

\subsection{Dependency Resolution}

Dependency resolution plays a big role in the success or failure of
building projects. Various other works focus on this topic and have
employed various techniques to mine open source repositories and
obtain project dependency information.

The work closest to ours is that presented by
Ossher~\cite{ossher-thesis} (Chapter 4). Like in our work, the
approach presented there also relies on the collection of a very large
number of JAR files, and on the creation of an index that maps FQNs to
JAR files, which is then used to find the dependencies for the
projects. Ossher's work had a different purpose than ours: they
produced a map of project-to-project dependencies across the Java
ecosystem, and were not trying to compile the projects. This, in turn,
carries a couple of technical differences between our work and
theirs. First, the JARs used by Ossher were a combination of those
present in the Maven Central Repository as well as a relatively
complicated system for identifying {\em projects} inside JAR
files. Instead of doing that, we simply collect all JAR files in all
the projects, and we use the ones that match the needed FQNs,
independent of what else they contain that may not be relevant for the
projects. Our approach is much simpler, and seems to work well for
purposes of compilation. Second, Ossher's dependency resolution did
not give priority to the JARs present in the projects, and looked for
matches in the entire collection of JARs. We started by doing the
same, but soon realized that the results of compilation were much
better if we gave priority to those local JARs.

Lungu et al. worked on extracting inter project dependencies with a
specific focus on Smalltalk ecosystems
\cite{Lungu:2010:RID:1858996.1859058}. Zhang, Hanyu details four
techniques for dependency resolution in C/C++ projects
\cite{zhang2011comprehensive}, through a) analyzing the source code of
the project; b) parsing build scripts; c) reading binary files with
the \emph{ELF} file format and discovering dynamically linked
libraries; and d) using a specification based technique that looks at
the source specifications of projects recorded in \emph{Debian}
repositories. Clearly, each programming language has its own
characteristics regarding dependencies; Smalltalk, C/C++ and Java are
all quite different. The idea of parsing build scripts in search for
dependencies, however, is one that we plan to explore in the future as
a way of improving JBF.

API usage analysis, like \cite{Xie:2006:MMA:1137983.1137997}, can be
used to design alternative techniques for finding the right
dependencies for projects. L\~{a}mmel et al. extracted API usage and
package footprint from large corpus of built java projects using
\emph{AST} based approaches. Here, missing packages required to build
the java projects were obtained by searching the web for the jars
using the unresolved package names as the search queries. They were
able to improve their build success rate by 15\% by using the
resolution method in \cite{Lammel:2011:LAA:1982185.1982471}.

\subsection{Empirical Studies of Builds}

An analysis of build systems in multi-language software can be found
in the work by Neitsch \textit{et al.} \cite{6405265} where the
authors performed a qualitative study on a set of five multi-language
open source software packages. They found that many build problems can
be systematically addressed, and make the general point that build
quality of software is rarely the subject of research, even though it
is paramount to maintaining and reusing software, a message we endorse.

Sulir \textit{et al.} \cite{Sulir:2016:QSJ:3001878.3001882} report on
work similar to ours. They attempted to build a collection of 7,264
Java projects from GitHub, from which around 60\% succeeded. This is a
higher rate than that obatined by JBF in the entire dataset, but it is
close to the rate we obtained for the smaller dataset of projects that
have build scripts, 56\%. Indeed, they only build projects which
contain build scripts from \ant, \maven or \gradle. Additionally,
their criteria to sample projects was much stricter than ours: at
least one fork, and the enforcement of a black list for specific
API's. We had no such constraints. Similar to what we observed, most
failures were due to missing dependencies. They also observed that
projects that built were smaller and older on average, and that the
number of stars does not seem to have any correlation with building
success. We did not analyze meta information related to age or other
social and GitHub aspects, but their conclusion regarding the impact
of scale goes in the same direction as our comparison between the
projects built in the first, and in the second round of compilation.

Seo \textit{et al.} \cite{Seo:2014:PBE:2568225.2568255} present a
large-scale empirical study of 26.6 million builds produced during a
period of nine months by thousands of developers at Google. They
analyze failure frequency, compiler error types, and resolution
efforts to fix those compiler errors. Two languages are analyzed, C++
and Java, and 37.4\% and 29.7\%, respectively, of these builds
fail. The most common errors are associated with dependencies between
components. The build environment is a Google's proprietary
cloud-based build process that utilizes a proprietary build language,
which is different from our generic building framework, but problems
of external dependencies were nevertheless similar to ours.

On the subject of recovering type information for Java programs,
Dagenais \textit{et al.} \cite{Dagenais:2008:ESA:1449764.1449790}
present a framework with an heuristics-based type inference system to
recover declared types of expressions and resolve ambiguities in
partial programs obtained from the Web. Their technique relies on the
generation of ASTs for each file. They test with 4 projects, three
'self-contained' and the fourth with more than 90 external
dependencies. They report around 90\% success among the types the
framework was able to recover, but between 35\% and 50\% of the types
remained unknown. Not surprisingly, the projects that
contained more external dependencies were on the higher range of
unknown types. JBF does not explicitly generate ASTs, relying instead
on javac to do that work.

\section{Conclusion}
\label{sec:conclusion}
We have presented \jbf, a tool to build \java programs on scale, with
automatic error repair and management of external dependencies.

Under direct comparison, the tool performed similarly to standard \java build
frameworks, building less 10\% of the projects. However, on a corpus
of \projectCount \java projects, \jbf build 24\% more projects than
what was possible with existing build frameworks, reaching 54\% of the
total projects built while introducing assurances of security, performance and integrity.

Two details about the results we obtain suggest there is a good potential to increase the success ratios of \jbf. First, the errors obtained after the second round of compilation were mostly due to unknown types, one of them explicitly created by \jbf signaling this exact problem. Second, the build framework that has the capability of obtaining external dependencies, \maven, was the one that outperformed \jbf. These suggest that an important threshold on success ratios had its origin on the availability of \jar files rather than any idiosyncrasy of the tool.

Increasing the size of the \jar corpus is ongoing work, either manually or through automation. For example, a new task of rating missing dependencies by frequency, after the second round of compilation, associating them with \jar files, obtain them on online sources and put them in our central repository is operationally simple\footnote{The Maven Central Repository, for example, allows searching \jar files by Classname, aka FQN.} and will further improve the built projects without affecting the \jbf pipeline.

Attached to the introduction of \jbf we created 50K-C, a dataset of \java programs together with
all their dependencies, with individual building scripts and with the class files generated by them.

\jbf  and 50K-C can benefit and accelerate research projects that target: the actual process
of large-scale building and running software, improvements on the \java programming language, and general domains of language engineering and understanding.

\bibliographystyle{abbrv}
\bibliography{bib/ase,bib/stats,bib/bib}

\end{document}